# Computing the coefficients for the power series solution of the Lane-Emden equation with the Python library SymPy

Klaus Rohe, D-85625 Glonn, email: klaus-rohe@t-online.de

## Abstract

It is shown how the Python library Sympy can be used to compute symbolically the coefficients of the power series solution of the Lane-Emden equation (LEE). Sympy is an open source Python library for symbolic mathematics. The power series solutions are compared to the numerically computed solutions using matplotlib. The results of a run time measurement of the implemented algorithm are discussed at the end.

## Derivation of the recurrence formulas for the coefficients of the power series solution of the LEE

In stellar physics the LEE is a second order nonlinear ordinary differential equation (ODE) which describes the internal state of a gaseous polytrophic sphere of index $n$, for details see reference (1), chapter 4. The LEE reads as follows:

**(1)** $\frac{1}{x^2} * \frac{d}{dx}\left(x^2 * \frac{d}{dx}f(x)\right) = -[f(x)]^n$ or

**(2)** $\frac{d^2}{dx^2}f(x) + \frac{2}{x}\frac{d}{dx}f(x) + [f(x)]^n = 0$

We are looking for a power series solutions $f(x) = \sum_{k=0}^{\infty} a_k * x^k$ of the LEE which satisfy the initial conditions $f(0) = 1$ and $\left[\frac{d}{dx}f(x)\right]_{x=0} = 0$. The first and second derivatives of the power series as well as its n[th] power have to be computed:

[a] $\frac{d}{dx}f(x) = \sum_{k=0}^{\infty} k * a_k * x^{k-1}$

[b] $\frac{2}{x}\frac{d}{dx}f(x) = 2 * \sum_{k=2}^{\infty} k * a_k * x^{k-2}$  ($a_1 = 0$ see below)

[c] $\frac{d^2}{dx^2}f(x) = \sum_{k=2}^{\infty} k * (k-1) * a_k * x^{k-2}$

[d] $[f(x)]^n = \left[\sum_{k=0}^{\infty} a_k * x^k\right]^n = \sum_{k=0}^{\infty} c_k * x^k$ for the $c_k$ the following recurrence formulas hold: $c_0 = a_0^n$ and $c_k = \frac{1}{k*a_0} * \sum_{l=1}^{k}[l*(n+1)-k] * a_l * c_{k-l}$ (see reference (3) or http://en.wikipedia.org/wiki/Formal_power_series)

From the initial conditions $f(0) = 1$ and $\left[\frac{d}{dx}f(x)\right]_{x=0} = 0$ it follows that $a_0 = 1$ and $a_1 = 0$.

Inserting [b] – [d] into equation (2) on gets:

$$\sum_{k=2}^{\infty} k * (k-1) * a_k * x^{k-2} + 2 * \sum_{k=2}^{\infty} k * a_k * x^{k-2} + \sum_{k=0}^{\infty} c_k * x^k = 0$$

From the above equation on gets by an index shift $k \rightarrow k-2$ on the third sum on the left side the following:

$$\sum_{k=2}^{\infty} k * (k-1) * a_k * x^{k-2} + 2 * \sum_{k=2}^{\infty} k * a_k * x^{k-2} + \sum_{k=2}^{\infty} c_{k-2} * x^{k-2} = 0$$

This can be further simplified to

$$\sum_{k=2}^{\infty}[k * (k-1) + 2 * k] * a_k * x^{k-2} + \sum_{k=2}^{\infty} c_{k-2} * x^{k-2} = 0 \text{ and finally one gets}$$

$$\sum_{k=2}^{\infty}\left[(k^2 + k) * a_k + c_{k-2}\right] * x^{k-2} = 0.$$

The last equation can be valid for arbitrary values of x only if for all $k \geq 2$ the following equation holds $(k^2 + k) * a_k + c_{k-2} = 0$. This finally results in the following recurrence formulas to compute the coefficients for the power series solution of the Lane-Emden equation:

**Box 1 Recurrence formulas for the coefficients of the power series solution of LEE**

For $k \geq 2$:

$$a_k = -\frac{c_{k-2}}{k^2 + k}$$

$$c_k = \frac{1}{k} * \sum_{l=1}^{k}[l * (n+1) - k] * a_l * c_{k-l}$$

For *k = 0, 1*: $a_0 = 1$ , $a_1 = 0$ and $c_0 = a_0^n = 1$, $c_1 = 0$ (see above).

The solutions of the LEE with the boundary values $f(0) = 1$ and $\left[\frac{d}{dx}f(x)\right]_{x=0} = 0$ are even functions (see reference (3), chapter 4) which means that all coefficients of the power series with odd indices are zero or more formally put $a_{2*k+1} = 0$, $k \in \{0, 1, 2, ..\}$.

## Computing $a_2, a_4, a_6$ and $a_8$ manually

For k = 2 we immediately get

$$a_2 = -\frac{c_0}{2^2 + 2} = -\frac{1}{2^2 + 2} = -\frac{1}{6}.$$

$$a_2 = -\frac{1}{6}.$$

**Case k = 4**

From the recurrence formula in the box above we get for $a_4$:

$$a_4 = -\frac{c_2}{4^2 + 4} = -\frac{c_2}{20}$$

So one has to compute $c_2$ with recurrence formula from the box above:

$$c_2 = \frac{1}{2} * \sum_{l=1}^{2} [l * (n + 1) - 2] * a_l * c_{2-l} = \frac{1}{2} \{ [1 * (n + 1) - 2] * a_1 * c_1 + [2 * (n + 1) - 2] * a_2 * c_0 \} = -\frac{n}{6}$$ so one gets

$$c_2 = -\frac{n}{6}$$

This finally gives the result

$$a_4 = \frac{n}{120}$$

## Case k = 6

The recurrence formula in the box above for $a_6$ yields:

$$a_6 = -\frac{c_4}{6^2 + 6} = -\frac{c_4}{42}$$

Computing $c_4$ with recurrence formula from the box above:

$$c_4 = \frac{1}{4} * \sum_{l=1}^{4} [l * (n + 1) - 4] * a_l * c_{4-l}$$
$$= \frac{1}{4} \{ [1 * (n + 1) - 4] * a_1 * c_3 + [2 * (n + 1) - 4] * a_2 * c_2 + [3 * (n + 1) - 4] * a_3 * c_1 + [4 * (n + 1) - 4] * a_4 * c_0 \}$$

Because $a_1 = a_3 = 0$ this simplifies the above expressions for $c_4$:

$$c_4 = \frac{1}{4} \{ [2 * (n + 1) - 4] * a_2 * c_2 + [4 * (n + 1) - 4] * a_4 * c_0 \}$$

Putting the values for $a_2$, $a_4$ and $c_0$ into the above formula gives

$$c_4 = \frac{1}{4} \Big\{ [2 * (n + 1) - 4] * \Big( -\frac{1}{6} \Big) * \Big( -\frac{n}{6} \Big) + [4 * (n + 1) - 4] * \Big( \frac{n}{120} \Big) * 1 \Big\}$$
$$= \frac{1}{4} \Big\{ [2 * n - 2] * \Big( \frac{n}{36} \Big) + [4 * n] * \Big( \frac{n}{120} \Big) \Big\} = \frac{(n - 1) * n}{72} + \frac{n^2}{120}$$
$$= \frac{5 * (n - 1) * n + 3 * n^2}{360} = \frac{8 * n^2 - 5 * n}{360}$$

$$c_4 = \frac{8 * n^2 - 5 * n}{360}$$

For $a_6$ we finally get $a_6 = -\frac{1}{42} * \frac{8*n^2 - 5*n}{360} = -\frac{8*n^2 - 5*n}{15120}$

$$a_6 = -\frac{8 * n^2 - 5 * n}{15120}$$

## Case k = 8

$a_8$ results from the recurrence formula in the box above:

$$a_8 = -\frac{c_6}{8^2 + 8} = -\frac{c_6}{72}$$

$$c_6 = \frac{1}{6} * \sum_{l=1}^{6} [l * (n + 1) - 6] * a_l * c_{6-l}$$

$$= \frac{1}{6} * \{[1 * (n + 1) - 6] * a_1 * c_5 + [2 * (n + 1) - 6] * a_2 * c_4 + [3 * (n + 1) - 6] * a_3 * c_3 + [4 * (n + 1) - 6] * a_4 * c_2 + [5 * (n + 1) - 6] * a_5 * c_1 + [6 * (n + 1) - 6] * a_6 * c_0\}$$

Because $a_1 = a_3 = a_5 = 0$ the above expression for $c_6$ now becomes:

$$c_6 = \frac{1}{6} * \{[2 * (n + 1) - 6] * a_2 * c_4 + [4 * (n + 1) - 6] * a_4 * c_2 + [6 * (n + 1) - 6] * a_6 * c_0\}$$

Substituting the values for $a_2$, $a_4$, $a_6$, $c_0$, $c_2$ and $c_4$ on gets:

$$c_6 = \frac{1}{6} * \left\{ (2 * n - 4) * \left(-\frac{1}{6}\right) * \left(\frac{8*n^2 - 5*n}{360}\right) + (4 * n - 2) * \left(\frac{n}{120}\right) * \left(-\frac{n}{6}\right) + (6 * n) * \left(-\frac{8*n^2 - 5*n}{15120}\right) \right\} = -\frac{1}{6} * \left\{ \frac{(2*n - 4)*(8*n^2 - 5*n)}{2160} + \frac{(4*n - 2)*n^2}{720} + \frac{6*n*(8*n^2 - 5*n)}{15120} \right\} = -\frac{1}{6} * \left\{ \frac{(2*n - 4)*(8*n^2 - 5*n)}{2160} + \frac{(4*n - 2)*n^2}{720} + \frac{6*n*(8*n^2 - 5*n)}{15120} \right\} = -\frac{1}{6} * \left\{ \frac{16*n^3 - 42*n^2}{2160} + \frac{(4*n - 2)*n^2}{720} + \frac{48*n^3 - 30*n^2}{15120} \right\} = -\frac{122*n^3 - 183*n^2 + 70*n}{45360}$$

$$a_8 = -\frac{1}{72} * -\frac{122*n^3 - 183*n^2 + 70*n}{45360}$$

So for $a_8$ we get:

$$a_8 = \frac{122*n^3 - 183*n^2 + 70*n}{3265920}$$

So until now we have $a_0 = 1$, $a_2 = -\frac{1}{6}$, $a_4 = \frac{n}{120}$, $a_6 = -\frac{8*n^2 - 5*n}{1520}$ and $a_8 = \frac{122*n^3 - 183*n^2 + 70*n}{3265920}$ which means for the Maclaurin series of $f(x)$:

$$f(x) = 1 - \frac{1}{6} * x^2 + \frac{n}{120} * x^4 - \frac{8*n^2 - 5*n}{1520} * x^6 + \frac{122*n^3 - 183*n^2 + 70*n}{3265920} * x^8 + \cdots.$$

## Aside

We will see that all power series coefficients except $a_0$ and $a_2$ are functions of the index $n$ with the property $a_{2*k}(n = 0) = 0$ for $k > 2$. This means for index $n = 0$ and the initial conditions $f(0) = 1$ and $\left[\frac{d}{dx} f(x)\right]_{x=0} = 0$ the LEE has the solution $f(x) = 1 - \frac{1}{6} * x^2$ which can be easily verified by inserting $1 - \frac{1}{6} * x^2$ into equation (1) or (2).

If one looks at the first table of appendix C one can verify that the coefficients for the power series solution of the LEE with index $1$ follow the rule $a_{2k} = \frac{(-1)^k}{(2*k+1)!}$.

The power series $\sum_{k=0}^{\infty} \frac{(-1)^k}{(2*k+1)!} * x^{2*k}$ represents for all $x \in \mathbb{R}$ the function $\frac{\sin(x)}{x}$. Inserting this into equation (1) or (2) one finds that it is also a solution of the LEE with the required initial conditions.

It is easily seen now that the computation of the next coefficients using the recurrence formulas becomes more and more tedious and error prone. Help comes from computer algebra systems which can be programmed to evaluate the recurrence formulas stated in the box above automatically. Here we will use SymPy (reference (4)) a Python library for symbolic computing. SymPy together with scipy for numerical computing and matplotlib for the visualization makes up a powerful environment for scientific and technical computing in Python. All the tools mentioned here are available as open source.

## Computing the coefficients of the power series with SymPy

The computation of the coefficients has been automated by using the Python library SymPy for symbolic mathematics to implement a Python class called **LEEPowerSeriesCoefficients**, a listing of the source code is given in appendix A. The class computes the m first coefficients of the power series solution (Maclaurin series) of the LEE and provides some methods to get the polynomial of degree m as an approximation for the Maclaurin series as well as to print the coefficients as a function of the index n of the LEE and to evaluate the coefficients for a given n as rational numbers (fractions). The symbolic computation of the coefficients by the recurrence formulas stated in the box above takes place in the constructor of the class (__init__ method). The parameter m of the constructor determines how many coefficients are computed. It should be observed that only coefficients which are not equal to zero are evaluated and printed. In appendix B the first 15 non zero coefficients of the power series are listed as a function of the index n. Appendix C shows the first 15 non zero coefficients of the power series for index n = 1 and n = 3 as rational numbers. They have been computed using the method **computeLEEPowerSeriesCoefficientsOfIndex(…)** of the above described Python class. This method prints the coefficients to **stdout** and returns a Python dictionary which contains the values of the coefficients as fractions represented as strings. They keys of the dictionary are the indexes of the coefficients. The dictionary contains coefficients only which are not equal to zero. In appendix C on one sees that the numerator and denominator can become very large integers. These fractions can be evaluated with the Python Decimal type to get a higher precession than with the standard floating point arithmetic.

## Comparison of the power series solution of the LEE with the results of numerical integration

In the following three figures the power series[1] solutions of the LEE for index n = 1.5 and 3 are compared to the solution which is obtained by using a numerical integration procedure similar to that described in reference (2), page 109 -123. The Python source code for the numerical integration procedure is shown in appendix D. The figures have been generated with the Python library matplotlib, see reference (4).

---

[1] More precisely it is a Maclaurin series, see http://mathworld.wolfram.com/MaclaurinSeries.html

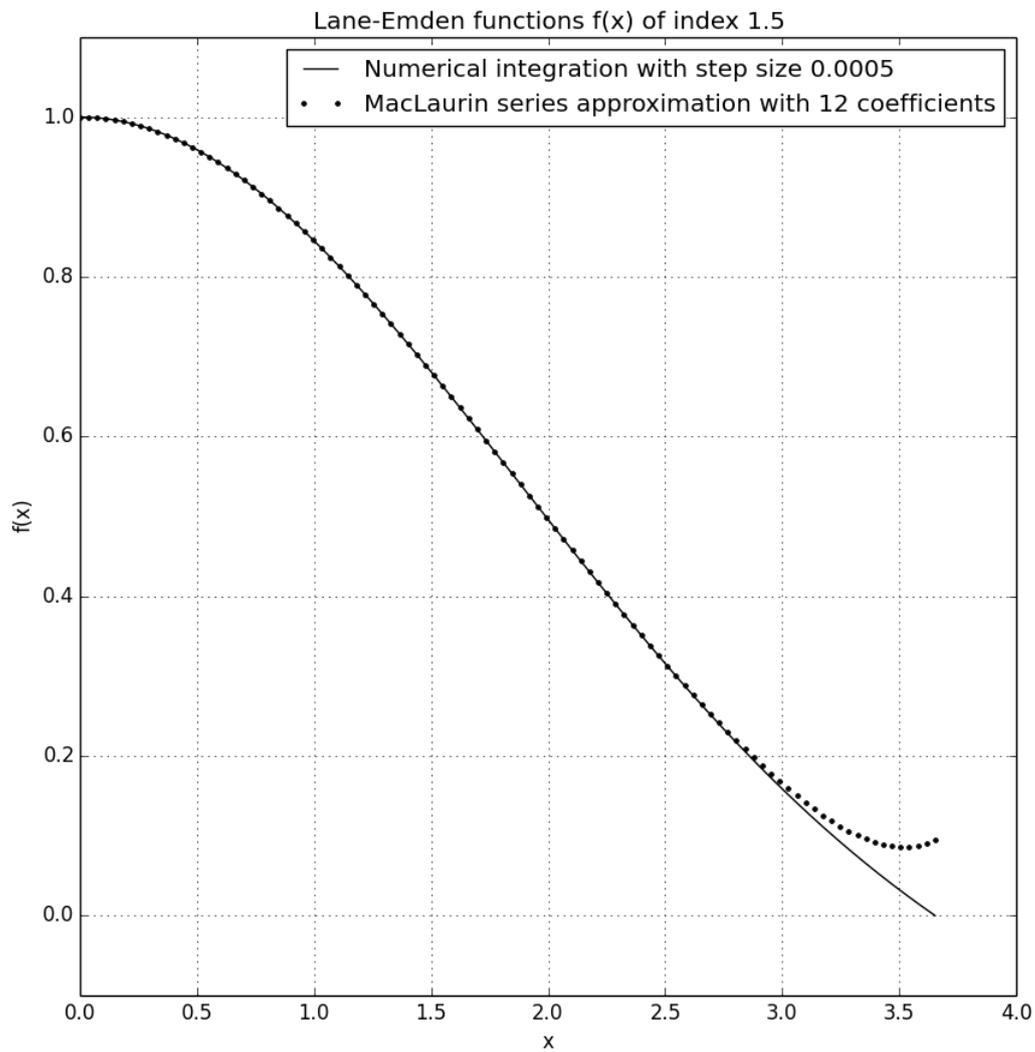

**Figure 1 Solution of the LEE with index 1.5 computed by numerical integration compared to the polynomial with the first 12 coefficients of the power series solution.**

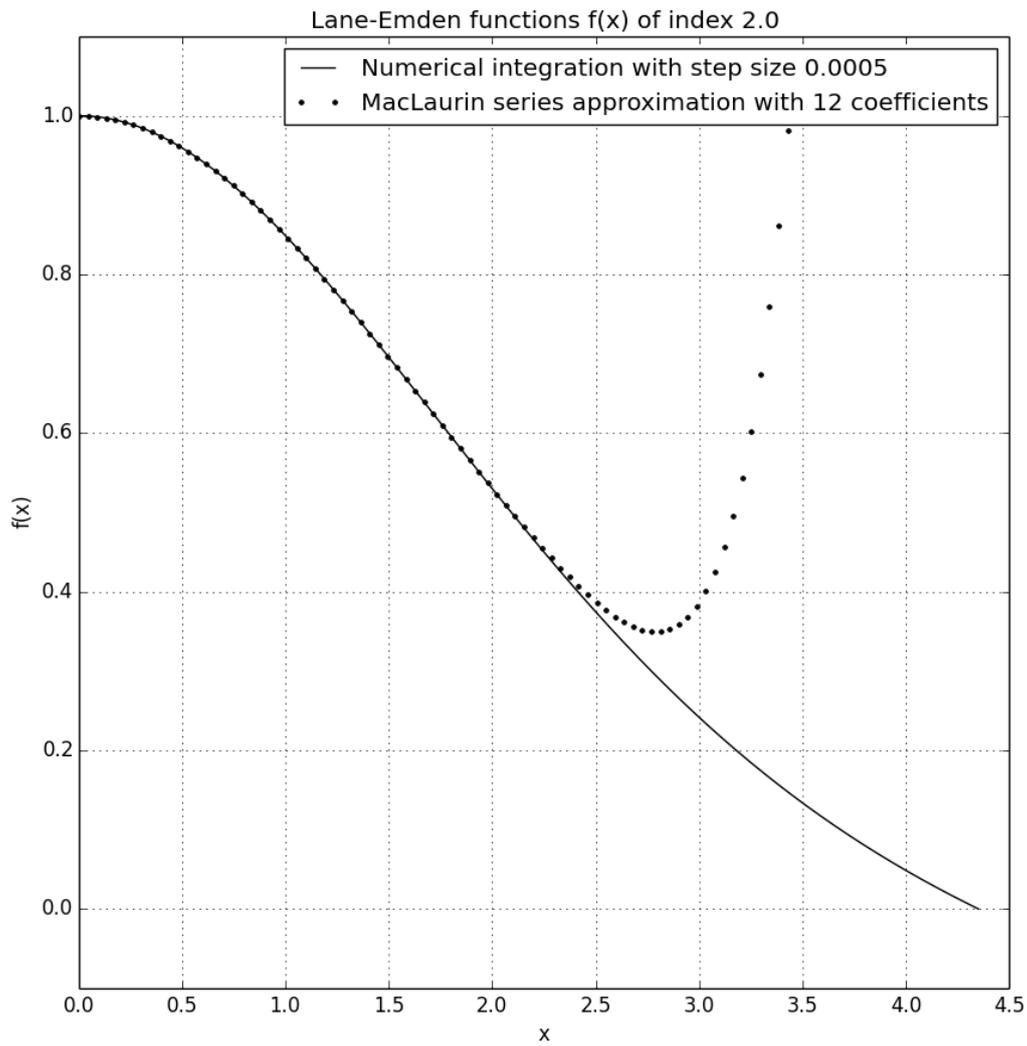

**Figure 2 Solution of the LEE with index 2 computed by numerical integration compared to the polynomial with the first 12 coefficients of the power series solution.**

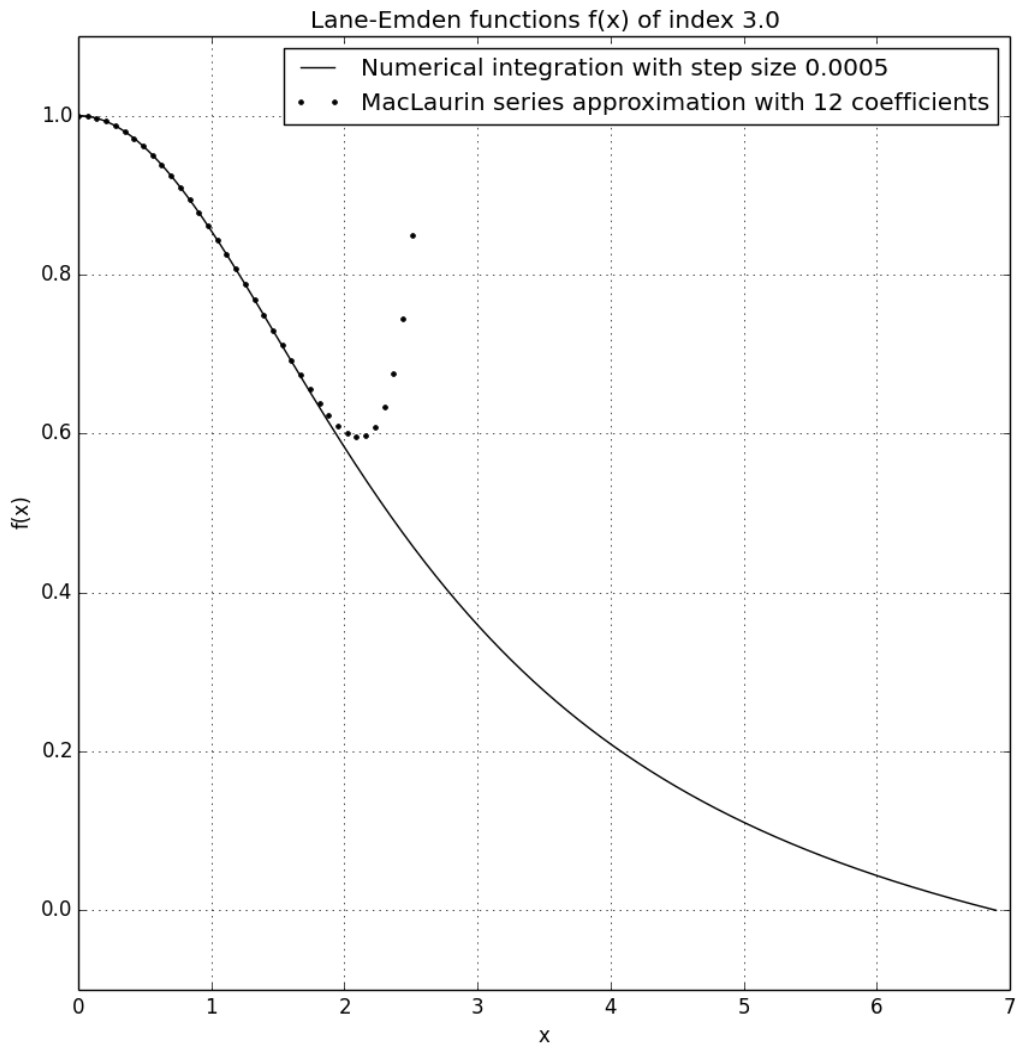

**Figure 3 Solution of the LEE with index 3.0 computed by numerical integration compared to the polynomial with the first 12 coefficients of the power series solution.**

## Run time analysis

The run time of the initialization of the Python class `LEEPowerSeriesCoefficient` has been measured as a function of the number of power series coefficients to compute (it has to be observed that all coefficients are computed during the initialization of the class). The environment for the measurement was Python 3.3.3 on 32 bit Windows 7 which runs on a computer with an Intel quad core i5-2320 CPU with 3.0 MHz clock frequency and 4 GB of main memory. The results are shown in figure 4. The logarithm of the elapsed time (run time) is plotted versus the number n of power series coefficients to compute. In the range $20 < n \leq 140$ the logarithm of the elapsed time is nearly a linear function of $n$ which means that the run time to compute the power series coefficients does nearly exponentially grow with $n$ for $20 > n$.

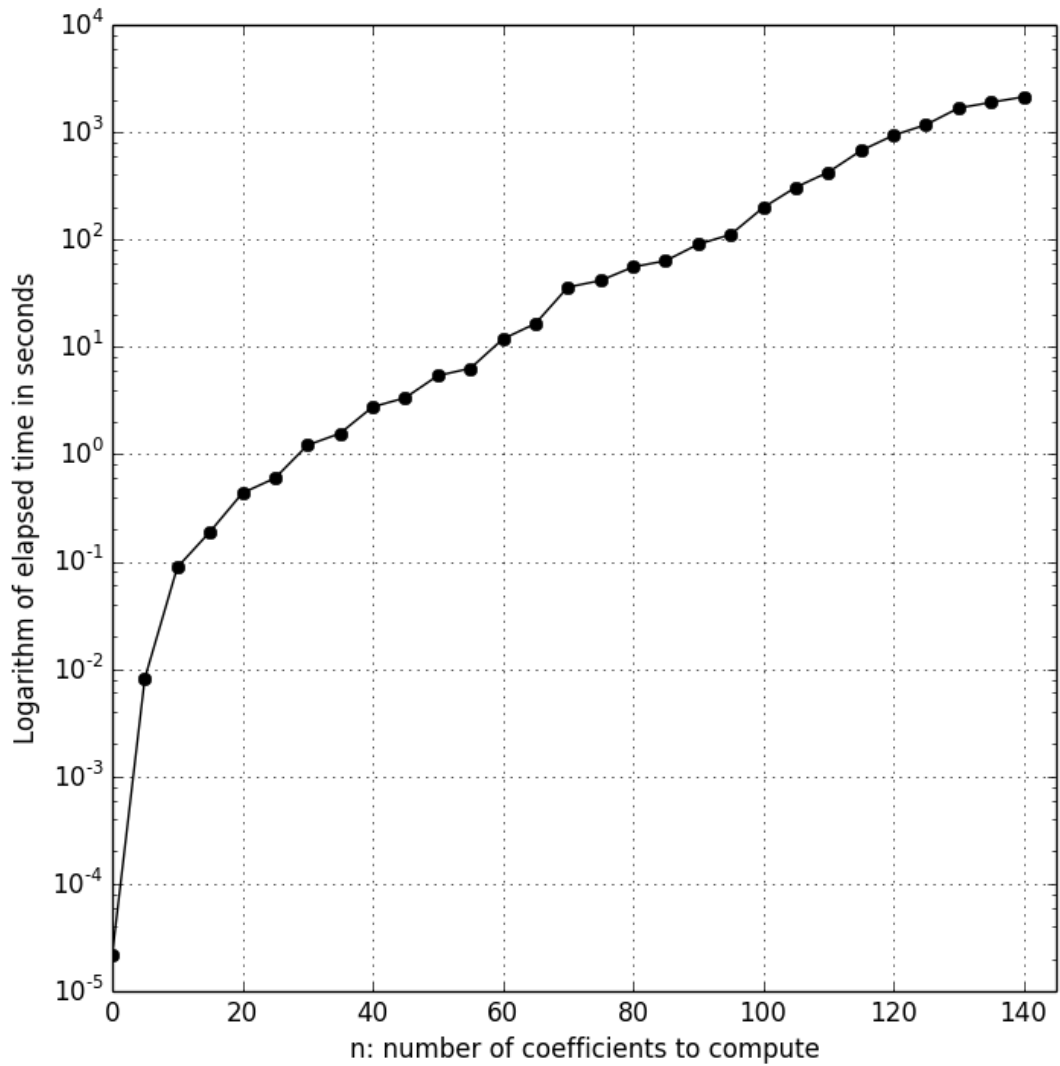

**Figure 4 Logarithm of elapsed time to compute n coefficients of the power series with SymPy**

# Appendices

## Appendix A:

## Source code of the Python class LEEPowerSeriesCoefficients

```python
# -*- coding: utf-8 -*-
"""
Created on Sun Aug 31 13:27:22 2014

Copyright (C) Klaus Rohe

The source code is provided as is and with no warranty of correctness.
"""

from sympy import *

class LEEPowerSeriesCoefficients:
    '''
    Computes the first m coefficients of the power series solution of the Lane-Emden equations
    (LEE) of index n using the symbolic math library SymPy.The following recursion formulas
    are used:

        for k = 0, 1:
            aks[0] = 1, aks[1] = 0, cks[0] = 0, cks[1] = 0

        for k >= 2:
            a[k] = -cks[k-2] / (k**2 + k)

                            k
            cks[k] = (1 / k) * \‾ (l * n + 1 - k) * aks[l] * cks[k - l]
                             /_

                            l=1

        'aks' is the array which contains the coefficients as SymPy expressions.
        'a' is a dictionary, which contains the coefficients as lambda expressions,
        so a[k] is a function of n.
    '''
    def __init__(self, m):
        """
        Constructor of the class which computes all coefficients
        of the power series up to degree m.
        """
        self.a = {0: lambda n: 1}
        # S(0), S(1) means that 0 and 1 are treated as symbols
        # and not as integers. See the documentation of SymPy for
        # keyword 'sympify'.
        self.aks = [S(1), S(0)]
        self.cks = [S(1), S(0)]
        n = Symbol('n')
        for k in range (2, m + 1):
            if k % 2 == 0:
                # Coefficients with even index are not equal to zero
                aexpr = sympify(- self.cks[k-2] / (S(k)**2 + S(k)))
                self.aks.append(factor(aexpr))
                cexpr = 0
                for l in range(1, k + 1):
                    cexpr = cexpr + (n * S(l) + S(1) - S(k)) * self.aks[l] * self.cks[k - l]
                self.cks.append(factor((S(1) / S(k)) * cexpr))
                self.a[k] = eval("lambda n: {0}".format(aexpr))
            else:
                self.aks.append(S(0))
                self.cks.append(S(0))

    def leeMacLaurinSeries(self, n):
        """
        Computes the polynom of degree m which approximates the MacLaurin series
        solution for the LEE of index n.
        """
        a_keys = sorted(list(self.a.keys()))
        def f(x):
            res = 0.0
            for k in a_keys:
                res = res + self.a[k](n) * x**k
            return res
        return f

    def computeLEEPowerSeriesCoefficientsOfIndex(self, n):
        """
        Evaluates the coefficients for a given n as rational numbers (fractions).
        It returns a dictionary which contains the coefficients as strings.
        For example a[8] = "619/1088640"
```

```python
        """
        a = {}
        print("\nCoefficients of the power series solution of LEE of index {0}".format(n))
        for i in range(0, len(self.aks)):
            if (i % 2 == 0):
                exprstr = "{0}".format(self.aks[i])
                exprstr = exprstr.replace("n", str(n))
                a[i] = "a[{0}] = {1}\n".format(i, S(exprstr))
                print(a[i])
        return a

    def getCoefficients(self):
        """
        Returns the dictionary 'a' which contains the coefficients
        as lambda expressions with parameter n and the array 'aks' which
        contains the coefficients as Sympy expressions.
        """
        return self.aks, self.a

    def writeCoefficients2File(self, fname):
        """
        Writes the array 'aks' to a file. Each line has the
        format i;expression where 'i' is the index of the coefficient
        and 'expression' the coefficient as function of n
        Example:
            000;1
            002;-1/6
            004;n/120
            006;-n*(8*n - 5)/15120
        """
        file = open(fname, "w")
        for i in range(0, len(self.aks)):
            if i % 2 == 0:
                file.write("{0:03d};{1}\n".format(i, self.aks[i]))
        file.close()

    def printCoefficients(self):
        """
        Writes 'aks' to stdout.
        """
        for i in range(0, len(self.aks)):
            if i % 2 == 0:
                print("a[{0}] = {1}\n".format(i, self.aks[i]))
```

## Appendix B:

## The 15 first non-zero power series coefficients as a function of n

The expressions for the 15 first nonzero power series coefficients haven been computed using the Python class 'LEEPowerSeriesCoefficients' listed above. The expressions for the coefficients in the box below are given in Python syntax. The operator ** means exponentiation.

```
a[0] = 1

a[2] = -1/6

a[4] = n/120

a[6] = -n*(8*n - 5)/15120

a[8] = n*(122*n**2 - 183*n + 70)/3265920

a[10] = -n*(5032*n**3 - 12642*n**2 + 10805*n - 3150)/1796256000

a[12] = n*(183616*n**4 - 663166*n**3 + 915935*n**2 - 574850*n +
138600)/840647808000

a[14] = -n*(21625216*n**5 - 103178392*n**4 + 200573786*n**3 - 199037015*n**2 +
101038350*n - 21021000)/1235752277760000

a[16] = n*(1442431856*n**6 - 8618115372*n**5 + 21827357636*n**4 - 30057075285*n**3
+ 23780949500*n**2 - 10267435500*n + 1891890000)/1008373858652160000

a[18] = -n*(368552598784*n**7 - 2657923739344*n**6 + 8348507232868*n**5 -
14830640277988*n**4 + 16120795594195*n**3 - 10740122081500*n**2 + 4066235428500*n
- 675404730000)/3103774736931348480000

a[20] = n*(65035924972928*n**8 - 551199819782480*n**7 + 2074925918891156*n**6 -
4538114873629364*n**5 + 6317195348852735*n**4 - 5740042719521900*n**3 +
3329284073314500*n**2 - 1128186384570000*n +
171102531600000)/6517926947555831808000000

a[22] = -n*(30720693974199296*n**9 - 299840088682556928*n**8 +
1319254687791147504*n**7 - 3438918097710593080*n**6 + 5860922969087284308*n**5 -
6782008348777403475*n**4 + 5335484162711174500*n**3 - 2754994980587692500*n**2 +
847953056599110000*n - 118574054398800000)/362787813900957598433280000000

a[24] = n*(4731477379473053696*n**10 - 52342890902954850528*n**9 +
264081052577164986584*n**8 - 801059938682391176900*n**7 +
1620103707989338077938*n**6 - 2285217511971127632065*n**5 +
2279636465710370388750*n**4 - 1589853990586539282500*n**3 +
742585473289204545000*n**2 - 209899877314257900000*n +
27272032511724000000)/6530180650217236771799040000000

a[26] = -n*(3348164045923792133427* n**11 - 414501122965955020208256*n**10 -
236214937631736974481548* n**9 - 818791094783083311544860*n**8 +
1920222289168048579662676*n**7 - 3202709967903276083371130*n**6 +
3880898046001433977673387*n**5 - 3420298046323193468157 3750*n**4 +
2150616934732129629242750*n**3 - 919574663361079656892500*n**2 +
240772767956779108050000*n -
29249254868823990000000)/5363498575249425250149423513600000000
```

## Appendix C:

## The 15 first non-zero power series coefficients for index 1 and 3

| Power series coefficients for index 1 |
|---|
| **a[0]** = 1 |
| **a[2]** = –1/6 |
| **a[4]** = 1/120 |
| **a[6]** = –1/5040 |
| **a[8]** = 1/362880 |
| **a[10]** = –1/39916800 |
| **a[12]** = 1/6227020800 |
| **a[14]** = –1/1307674368000 |
| **a[16]** = 1/355687428096000 |
| **a[18]** = –1/121645100408832000 |
| **a[20]** = 1/51090942171709440000 |
| **a[22]** = –1/25852016738884976640000 |
| **a[24]** = 1/15511210043330985984000000 |
| **a[26]** = –1/10888869450418352160768000000 |
| **a[28]** = 1/8841761993739701954543616000000 |

| Power series coefficients for index 3 |
|---|
| **a[0]** = 1 |
| **a[2]** = –1/6 |
| **a[4]** = 1/40 |
| **a[6]** = –19/5040 |
| **a[8]** = 619/1088640 |
| **a[10]** = –17117/199584000 |
| **a[12]** = 1208293/93405312000 |
| **a[14]** = –24355481/12482346240000 |
| **a[16]** = 407094043/1383228887040000 |
| **a[18]** = –463911176707/10450419989667840000 |
| **a[20]** = 107759617263073/1609364678408473600000 |
| **a[22]** = –452344982719313191/447886190001182220288000000 |
| **a[24]** = 122812575931580523743/806195142002127996518400000000 |
| **a[26]** = –8949885243979365830917­9/389506069371781063917­8956800000000 |
| **a[28]** = 4348102629052610323474­74509/125457308237521353480861­412556800000000 |

## Appendix D:

## Source code for the numerical solution of the LEE

```python
# -*- coding: utf-8 -*-
"""
Created on Wed Aug 20 19:25:18 2014

Copyright (C) Klaus Rohe

The source code is provided as is and with no warranty of correctness.
"""

def solveLaneEmdenByMidPointMethod(dx, n):
    """
    See:
    Paul Hennings, ASTROPHYSICS WITH A PC,
    An Introduction To Computational Astrophysics,
    Willmann-Bell, Richmond 1994, page 109 - 125

    Lane-Emden equation (LEE): d²F(x)/dx² + (2 / x) * dF(x)/dx + F(x)**n = 0

    With H(x) = dF(x)/dx the second order Lane-Emden ODE can be written as
    as system of two first order ODEs:

        dF(x)/dx = H(x)
        dH(x)/dx = -F(x)**n - (2 / x) * H(x)
    """
    xarr = [0.0]
    Farr = [1.0]
    Harr = [0.0]

    # Approximations of F(x) and H(x) by power series for small x
    F = lambda n, x: 1.0 - x**2 / 6.0 + (n / 120.0) * x**4 - (n * (8.0 * n - 5.0) / 15120.0)
* x**6 \
    + (n * (122 * n**2 - 183 * n + 70) / 3265920) * x**8 \
    - (n * (5032*n**3 - 12642*n**2 + 10805*n - 3150 )/1796256000) * x**10

    H = lambda n, x: - x / 3.0 + (n / 30.0) * x**3 - (n * (8.0 * n - 5.0) / 2520.0) * x**5 \
    + (n * (122 * n**2 - 183 * n + 70) / 408240) * x**7 \
    - (n * (5032*n**3 - 12642*n**2 + 10805*n - 3150 )/179625600) * x**9

    i = 0
    x = dx
    while True:
        if x <= 3.0 * dx:
            Farr.append(F(n, x))
            Harr.append(H(n, x))
        else:
            Fi1_2 = Farr[i] + 0.5 * dx * Harr[i]
            Hi1_2 = Harr[i] + 0.5 * dx * (-Farr[i]**n - (2.0 / xarr[i]) * Harr[i])
            Fi1 = Farr[i] + dx * Hi1_2
            xi1_2 = xarr[i] + 0.5 * dx
            Hi1 = Harr[i] + dx * (-Fi1_2**n - (2.0 / xi1_2) * Hi1_2)
            if Fi1 < 0.0:
                break
            Farr.append(Fi1)
            Harr.append(Hi1)
        x = x + dx
        xarr.append(x)
        i = i + 1
    return xarr, Farr, Harr
```